\documentclass[11pt]{iopart}
\usepackage{iopams,mathrsfs,graphicx,color}

\begin{document}

\title[Spin ladders and Luttinger liquids ]{Spin ladders and quantum simulators for Luttinger liquids}

\author{S. Ward$^{1,2}$, P. Bouillot$^3$, H. Ryll$^4$, K. Kiefer$^4$, K.W. Kr\"amer$^5$, Ch. R\"uegg$^{1,2}$, C. Kollath$^6$ and T. Giamarchi$^7$}

\address{$^1$ Laboratory for Neutron Scattering, Paul Scherrer Institut, CH--5232 Villigen PSI, Switzerland}
\address{$^2$ London Centre for Nanotechnology and Department of Physics and Astronomy,
University College London, London WC1E 6BT, United Kingdom}

\address{$^3$ Department of Neurosurgery, Geneva University Hospital, 1211 Geneva, Switzerland}

\address{$^4$ Helmholtz Center Berlin for Materials and Energy, D--14109 Berlin, Germany}

\address{$^5$ Department of Chemistry and Biochemistry, University of Bern, CH--3012 Bern, Switzerland}

\address{$^6$ DPT, University of Geneva, CH--1211 Geneva, Switzerland}

\address{$^7$ DPMC-MaNEP, University of Geneva, CH--1211 Geneva, Switzerland}
\ead{Thierry.Giamarchi@unige.ch}

\begin{abstract}
Magnetic insulators have proven to be usable as quantum simulators for itinerant interacting quantum systems.
In particular the compound (C$_{5}$H$_{12}$N)$_{2}$CuBr$_{4}$ (short (Hpip)$_{2}$CuBr$_{4}$) was shown to be a
remarkable realization of a Tomonaga-Luttinger liquid (TLL) and allowed to quantitatively test the TLL theory.
Substitution weakly disorders this class of compounds and allows thus to use them to tackle questions pertaining
to the effect of disorder in TLL as well, such as the formation of the Bose glass. As a first step in this direction
we present in this paper a study of the properties of the related (Hpip)$_{2}$CuCl$_{4}$ compound. We determine the
exchange couplings and compute the temperature and magnetic field dependence of the specific heat, using a finite temperature
Density Matrix Renormalization group (DMRG) procedure. Comparison with the measured specific heat at zero magnetic field
confirms the exchange parameters and Hamiltonian for the (Hpip)$_{2}$CuCl$_{4}$ compound, giving the basis needed to start
studying the disorder effects.
\end{abstract}





\section{Introduction}

Understanding the physics of quantum interacting systems is one of the most challenging problems of condensed matter physics.
This is specially true in low dimensions where the interaction effects are usually reinforced by the dimensional confinement and lead to novel physics. This is the case for one dimensional
quantum systems, for which the interactions lead to a set of properties quite different than their higher dimensional counterparts. In particular, all the excitations become collective ones and the usual Landau or Bogoliubov quasiparticles \cite{mahan_book} do not exist. The resulting physics, which presents a unique set of universal properties can be described by the Tomonaga Luttinger liquid (TLL) theory, characterized by powerlaw decay of the correlation functions and fractionalization of the excitations \cite{giamarchi_book_1d}.

TLL physics manifests itself in many different experimental situations, which are the object of this special issue,
and in particular the powerlaw behavior of the correlations has been observed in several situations \cite{schwartz_electrodynamics,bockrath_luttinger_nanotubes,lee_optics_chains_ybco,denlinger_arpes_LiMoO} (see also \cite{giamarchi_book_1d,cazalilla_rpm_bosons} for additional references).
However in many of the experimental systems, it is difficult to go beyond the very observation of the powerlaw behavior, and fully test the TLL theory. Indeed, in the condensed matter context, the interactions are usually poorly known (typically a screened coulomb interaction), which makes an ab-initio calculation of the Luttinger liquid exponent impossible. In addition it is usually difficult to test the predicted universality of the TLL behavior (namely that all the exponents are functionally related to a \emph{unique} parameter $K$ \cite{giamarchi_book_1d})
since one often lacks a control parameter or the possibility to easily access several correlation
functions simultaneously.

As a result, it is highly desirable to dispose of ``quantum simulators'' to realize TLLs, namely experimental systems that can be faithfully described by a simple model Hamiltonian and for which \emph{quantitative} comparison with the experiments is possible.
A prime candidate for such quantum simulators is provided by cold atomic systems \cite{bloch_cold_atoms_optical_lattices_review}.
Their remarkable versatility and degree of control of the dimensionality of the problem,
of the kinetic energy and interactions has made them invaluable tools
to tackle several properties of strongly correlated systems.
However for TLL physics, and despite remarkable success in realizing and probing one dimensional systems,
they still suffer from limitations coming from either the confining potential, that corresponds to a space varying chemical potential
and thus blurs the exponents, or limitations of interactions for the systems without the confining potential (see e.g. \cite{paredes_tonks_experiment,hofferberth_interferences_atomchip_LL}).
Another class of quantum simulators that has proven very successful are the magnetic insulators \cite{giamarchi_BEC_dimers_review}.
Indeed such spin systems can be mapped onto interacting boson systems. The density of bosons can be controlled
by a magnetic field and easily measured from the magnetization.
In high dimension, they have proven very successful to study effects such as the Bose-Einstein condensation \cite{giamarchi_coupled_ladders,nikuni_bec_tlcucl,ruegg_bec_tlcucl}.
In one dimension these systems provide remarkable realization
of a TLL. Because the ``interactions'' between the ``bosons'' are now provided by the magnetic exchanges,
they are very well known and short range, allowing a direct calculation
of the TLL parameters without any fudging. Comparison between such calculations and experimental results on the metal-organic spin ladder \cite{watson_bpcb} piperidinium copper bromide (C$_{5}$H$_{12}$N)$_{2}$CuBr$_{4}$, short (Hpip)$_{2}$CuBr$_{4}$, have provided the first \emph{quantitative} test
of the TLL theory \cite{klanjsek_nmr_ladder_luttinger,ruegg_thermo_ladder,thielemann_neutron_ladder,bouillot_dynamics_ladder_DMRG_long}.
Since these first experiments the successful comparison between theory and experiment in this class of compounds has been extended to more refined correlations such as the ones measured in Electron Spin Resonnance (ESR, \cite{furuya_ESR_BPCB}), or neutron scattering
experiments \cite{bouillot_dynamics_ladder_DMRG_long,thielemann_spectrum_spinladder}. Other classes of ladder compounds have also been investigated \cite{Hong2010,Schmidiger2011,schmidiger_dimpy_neutrons}.

One of the interesting effects that could be studied with such a class of material, are the effects of disorder on interacting quantum particles. Indeed, disordered bosons lead to interesting phases such as the Bose glass phase, predicted a long time ago \cite{giamarchi_loc,fisher_boson_loc}, whose experimental realization is still elusive. Magnetic insulators provide a very
nice playground for such effects since chemical substitution allows to weakly affect the spin exchange and thus to introduce a weak disorder on the bosons. Several materials
have been exploited to test for the presence of a Bose glass phase in various dimensions
ranging from three- to quasi-one dimensional systems \cite{giamarchi_BEC_dimers_review,hong_disorder_magnets,yamada_disorder_magnets,yu_disorder_magnets}. Given the importance of the (Hpip)$_{2}$CuBr$_{4}$ compound and the excellent realization of one dimensional TLL that it offers it is particularly important to be able to control the disorder effects in this material.

We thus introduce in this paper the disordered version (Hpip)$_{2}$CuBr$_{4(1-x)}$Cl$_{4x}$, in which the Br atoms have been replaced partially by Cl, and studies of the parent compounds (Hpip)$_{2}$CuBr$_{4}$ and (Hpip)$_{2}$CuCl$_{4}$. We investigate various quantities such as the specific heat and show how one can extract the various characteristic parameters of the system.

The plan of the paper is as follows. In Sec.~\ref{sec:BPCB} we present the general properties of the compounds and their theoretical description. In Sec.~\ref{sec:results} we present the results for the specific heat measurements and the comparison with the Density Matrix Renormalization Group (DMRG) calculations, as well as a description of the methodology. In Sec.~\ref{sec:conclusion} we discuss
some of the consequences and perspectives for this class of compound.

\section{Generalities on (Hpip)$_{2}$CuBr$_{4}$} \label{sec:BPCB}

\subsection{Compound} \label{sec:compound}

Recent successes in the synthesis and the growth of single crystals of new metal--organic compounds have opened up exciting new routes for experimental studies of model magnetic materials. This is due to the typical energy scale of the magnetic exchange interactions in such materials, which are on the order of meV, and exceptionally clean realizations of some low-dimensional exchange geometries, e.g. one--dimensional spin ladders or two--dimensional square--lattices.

\begin{figure}
 \begin{center}
  \includegraphics[width=0.5\columnwidth]{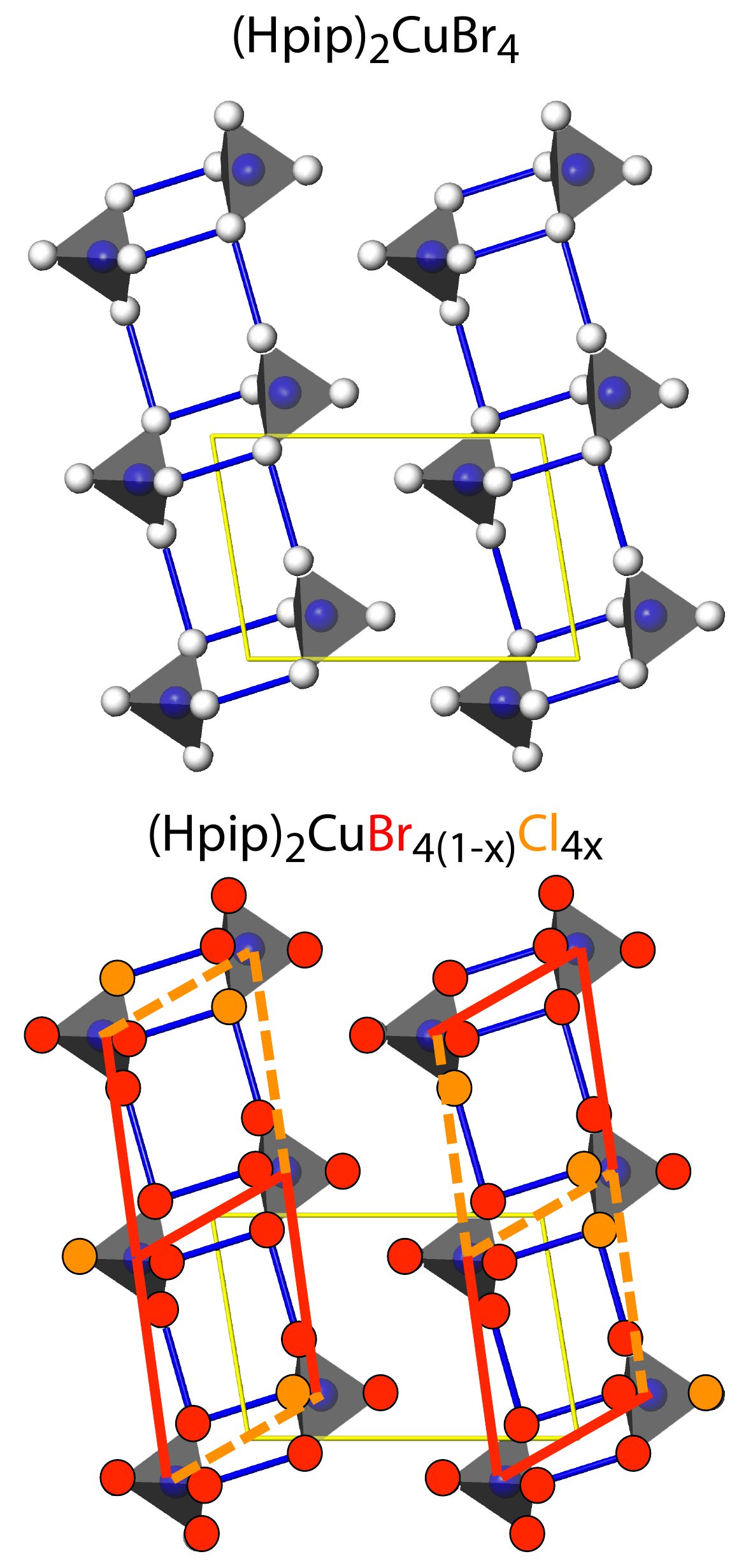}
 \end{center}
\caption{\label{fig:structure} Spin ladder units in (Hpip)$_{2}$CuBr$_{4}$. Top: Structure of the exchange paths mediated by Br$^{-}$ sites (white) between Cu$^{2+}$ ions carrying $S=1/2$ (blue). Bottom: Effect of partial substitution of Br$^{-}$ (red) by Cl$^{-}$ (orange) on the effective ladder exchange interactions. Dashed bonds will be affected by the substitution.}
\end{figure}

In (Hpip)$_{2}$CuBr$_{4}$ the magnetic Cu$^{2+}$ ions with quantum spin $S=1/2$ form one--dimensional ladder--like structural units. Magnetic exchange interactions in these ladder arrays are via Cu--Br--Br--Cu super--exchange paths. As demonstrated in Fig.~\ref{fig:structure}, two such paths contribute to the exchange on ladder rungs ($J_\perp$), whereas one, longer path constitutes the ladder leg ($J_\parallel$). Possible interladder exchange ($J'$) is very small due to the large organic (C$_{5}$H$_{12}$N)$+$ piperidinium ion effectively separating the ladder units.

While only experiments by neutron inelastic scattering are able to unambiguously determine the exchange Hamiltonian of such a spin system \cite{thielemann_spectrum_spinladder}, also measurements of bulk magnetic properties, such as the uniform magnetization with clear square--root field--dependencies near the critical magnetic fields, may indicate the excellent low--dimensionality of a material \cite{watson_bpcb, klanjsek_nmr_ladder_luttinger}. Once the ladder Hamiltonian is confirmed, the critical magnetic fields $h_{c1}$ and $h_{c2}$ are used to extract precise values for $J_\parallel$ and $J_\perp$ as will be described below.

The exchange parameters for (Hpip)$_{2}$CuBr$_{4}$, as summarized in Table~\ref{tab:exp}, place the material in the so--called strong--coupling limit of the quantum spin ladder. These critical fields allowed for the first time experimental studies of all its TLL properties by a number of high-precision experimental techniques \cite{watson_bpcb, klanjsek_nmr_ladder_luttinger,ruegg_thermo_ladder,thielemann_neutron_ladder,thielemann_spectrum_spinladder}.

\begin{table}
\caption{\label{tab:exp} Summary of ladder exchange parameters and critical fields from uniform magnetization data.}
\begin{tabular}{|l|c|c|c|c|}
\hline\hline
Compound & $J_\perp$(K) & $J_\parallel$(K) & $h_{c1}$(T) & $h_{c2}$(T) \\
\hline
(Hpip)$_{2}$CuBr$_{4}$ \cite{klanjsek_nmr_ladder_luttinger} & 12.6 & 3.55 & 6.73 & 13.79 \\
\hline
(Hpip)$_{2}$CuBr$_{2}$Cl$_{2}$ \cite{HpipClBr_JMMM} & 5.10 & 3.06 & 2.4 & 20 \\
\hline
(Hpip)$_{2}$CuCl$_{4}$ \cite{HpipClBr_JMMM} & 3.52 & 1.13 & 1.8 & 4.9 \\
\hline\hline
\end{tabular}
\end{table}

\begin{figure}
 \begin{center}
  \includegraphics[width=0.9\columnwidth]{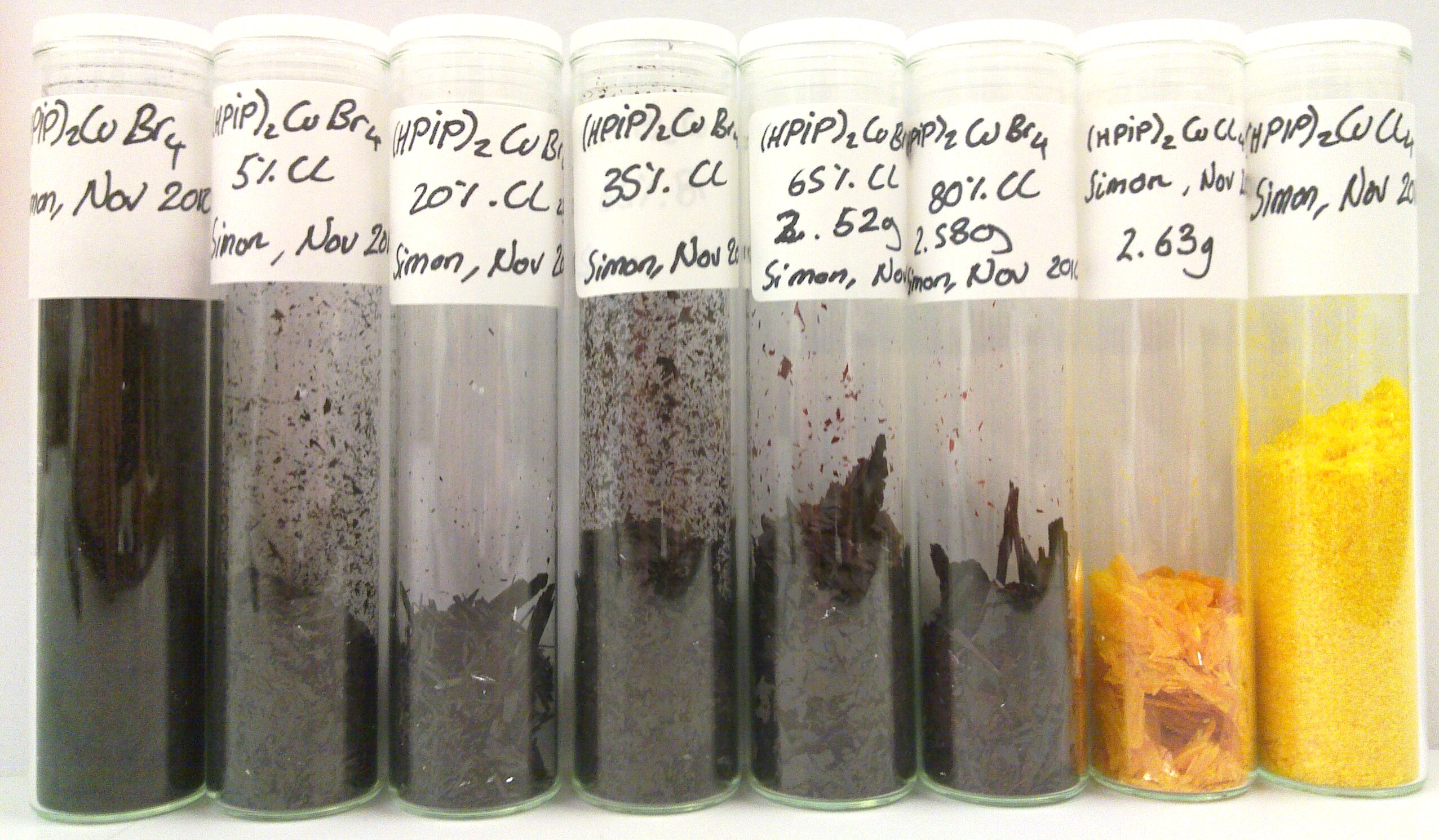}
 \end{center}
  \caption{\label{fig:HpipBrCl_chem} Complete Br/Cl substitution series (Hpip)$_{2}$CuBr$_{4(1-x)}$Cl$_{4x}$, x=0 (left, black) and x=1 (right, yellow).}
\end{figure}

In addition to the nearly optimal spin ladder properties of this compound its chemical flexibility can be explored to realize the full potential of such metal--organics as low--dimensional model systems, in which further aspects of TLL can be studied in great detail. Full and partial substitution of the Br by Cl has been demonstrated, see Fig.~\ref{fig:HpipBrCl_chem}. The chloride (Hpip)$_{2}$CuCl$_{4}$ is structurally identical to its bromine analog. Partial Br/Cl substitution in  (Hpip)$_{2}$CuBr$_{4(1-x)}$Cl$_{4x}$ takes place far from the Cu$^{2+}$ sites that carry the $S=1/2$ moments needed for the magnetic properties. It will thus not affect the general properties of the system and its ladder-like structure. It is thus much less violent than other forms of magnetic disorder, such as the replacement of the Cu$^{2+}$ ions by Zn $^{2+}$ ($S=0$), which fully removes a magnetic site. Since the Br/Cl--site affects the super--exchange paths, one can thus expect a modification of the value of the exchange parameters. Such modification is indicated in Fig.~\ref{fig:structure}. The parameters, extracted previously from magnetization data, for three compounds are summarized in Table~\ref{tab:exp}.

We will show in the following sections how one can characterize the compound (Hpip)$_{2}$CuCl$_{4}$, namely ascertain the structure of its Hamiltonian as well as determine its parameters.

\subsection{Theoretical description} \label{sec:theo}

The piperidinium compounds are very well described by a spin ladder Heisenberg Hamiltonian

\begin{equation} \label{equ:spinladderhamiltonian}
H_\mu= J_\perp H_\perp + J_\parallel H_\parallel
\end{equation}

where

\begin{eqnarray}
H_\perp=\sum_{l=1}^{L}\mathbf{S}_{l,1}\cdot\mathbf{S}_{l,2}-h^z J_\perp^{-1} M^z \label{equ:Hperp}\\
H_\parallel=\sum_{k=1,2}\sum_{l=1}^{L-1} \mathbf{S}_{l,k}\cdot\mathbf{S}_{l+1,k} \label{equ:Hparallel}
\end{eqnarray}

The magnetic field, $h^z,$ is applied in the $z$ direction, and $M^z$ is the $z$-component of the total spin operator $\mathbf{M}=\sum_{l=1}^{L}(\mathbf{S}_{l,1} +\mathbf{S}_{l,2})$. The operator $\mathbf{S}_{l,k}=(S_{l,k}^x,S_{l,k}^y,S_{l,k}^z)$ acts at the site $l$ ($l=1,2,\ldots,L$) of the leg $k$ ($k=1,2$). $S^\alpha_{l,k}$ ($\alpha=x,y,z$) are conventional spin-$1/2$ operators, $[S^x_{l,k},S^y_{l,k}]=iS^z_{l,k}$, and $S^\pm_{l,k}=S^x_{l,k}\pm iS^y_{l,k}$. In addition, there is a weak interladder coupling that we
will not consider here (for more details on its effects see Ref.~\cite{bouillot_dynamics_ladder_DMRG_long}). The magnetic field in
Tesla is related to $h^z$ of Eq.~\ref{equ:spinladderhamiltonian} by $\frac{h^z}{g\mu_B}$
with $\mu_B$ being the Bohr magneton and $g$ being the Land\'e factor
of the unpaired copper electron spins. ESR measurements have confirmed that anisotropies are only of the order of a few percents \cite{cizmar_esr_bpcb,furuya_ESR_BPCB}.

\begin{figure}
 \begin{center}
  \includegraphics[width=0.9\columnwidth]{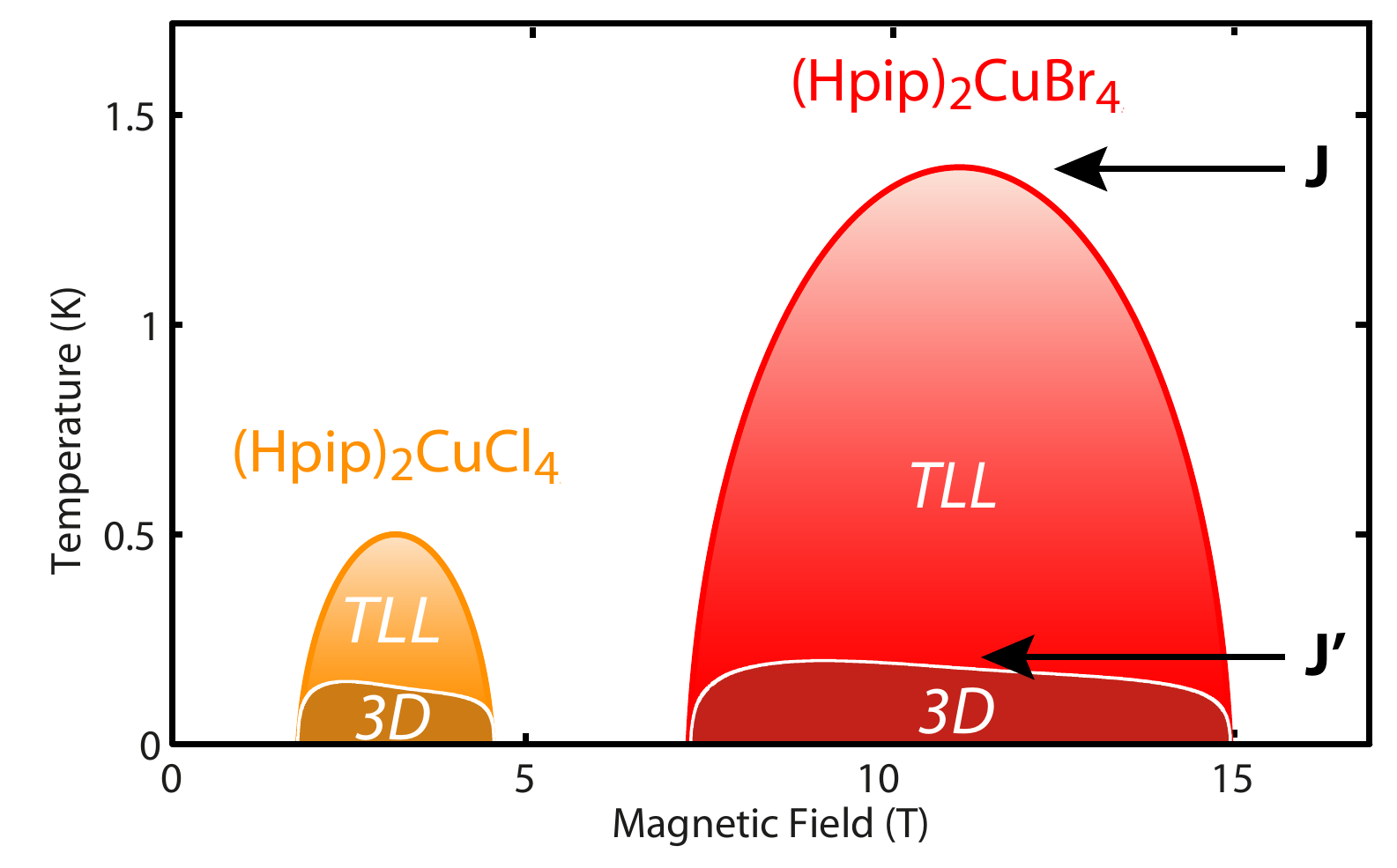}
 \end{center}
  \caption{\label{fig:phasediag} Schematic temperature-magnetic field phase diagram for the two compounds (Hpip)$_{2}$CuBr$_{4}$ and (Hpip)$_{2}$CuCl$_{4}$. The coherence scale (see text) of the order of the exchange $J = J_\parallel$ along the legs of the ladder is shown. Below this coherence scale the system is described by a TLL. If there is a weak interladder residual coupling $J'$ the spin will ultimately show three-dimensional antiferromagnetic planar order (3D).}
\end{figure}

In the absence of a magnetic field, the ground state of the system is a superposition of dimers, in a singlet state, separated by a gap of order $J_{\perp}$ from the triply degenerate excited triplet states. Application of the magnetic field reduces the energy of one of the triplets until it crosses the level of the singlet. Since the triplet can delocalize from rung to rung because of the exchange $J_{\parallel}$, we have a band of triplets. We thus have two quantum phase transitions. At $h_{c1}$ the first triplet (from the bottom of the band of triplets) enters the system while at $h_{c2}$ the triplet band is full. The triplets can be faithfully represented by hard core bosons, or in one dimension by spinless fermions using the Jordan-Wigner transformation. The magnetic field is thus acting as a ``gate voltage'' for an electronic system controlling the chemical potential of the triplets. The density of triplets is directly measured by the magnetization of the system which will increase from zero to $1$ per rung. The triplets are of course interacting due to the magnetic exchange. As indicated in Fig.~\ref{fig:phasediag}, the system has three phases. For $h < h_{c1}$ it is in a gapped state with essentially a ground state made of singlets. For $h_{c1} < h < h_{c2}$ it contains interacting triplets. It is thus a TLL. The excellent one dimensionality of the compound, the fact that the microscopic Hamiltonian is extremely well known and the control of the density of carriers makes it a system of choice to study the properties of TLL. For $h>h_{c2}$ the whole band of triplets has been filled and the system is gapped again.

Note that the two critical fields are directly related to the two coupling constants $J_\perp$ and $J_\parallel$ and thus give access to these parameters. For the (Hpip)$_{2}$CuBr$_{4}$ compound the parameters are indicated in Table~\ref{tab:exp}, and were determined using this technique and NMR measurements of the magnetization \cite{klanjsek_nmr_ladder_luttinger}.
In addition to the values of the parameters it is important to ascertain that no important term has been forgotten in the Hamiltonian (Dzialoshinski-Moriya interactions for example). This can be done by comparison of the specific heat with the results using the Hamiltonian (\ref{equ:spinladderhamiltonian}) and the coupling constants extracted from the magnetization.
This independent check provides a very stringent test for the theoretical description. For the (Hpip)$_{2}$CuBr$_{4}$ compound this program has been successfully carried out \cite{ruegg_thermo_ladder}.

We present in this paper a similar calculation of the specific heat for the chloride compound (Hpip)$_{2}$CuCl$_{4}$.

\section{Specific heat measurements} \label{sec:results}

\subsection{Theoretical methodology}

In order to make predictions that can be compared with experiments, we need to compute e.g. magnetization and specific heat for the Hamiltonian (\ref{equ:spinladderhamiltonian}).  The theoretical results are obtained using the variants of the density-matrix renormalization group method or also called matrix product state algorithms \cite{white_dmrg,Schollwock_DMRG,Hallberg_rev,Jeckelmann_rev}.
This method is a variational method, which relies on an optimization within the space of so-called matrix product states. The numerical method has been proven very powerful in particular describing the ground state or dynamic properties of one-dimensional systems at zero or finite temperature. Here we use the auxiliary state variant \cite{Verstraete_finiteT_DMRG,Zwolak_finiteT_DMRG,White_finT} in order to calculate thermodynamic properties as the magnetization or the specific heat at finite temperature.
In the calculations we obtained converged results for systems of $L = 60$ rungs, keeping $m=96$ states for the Hilbert space of each block.
For the temperature we kept an imaginary time step of $d\tau = 0.05$ 1/K.
The results for spin systems have been shown to be remarkably accurate (for more details see e.g. \cite{bouillot_dynamics_ladder_DMRG_long} and references therein). For the (Hpip)$_{2}$CuBr$_{4}$ compound an excellent agreement has been found between the measured and computed specific heat, confirming fully the form of the spin ladder Hamiltonian and the values of the exchange parameters \cite{ruegg_thermo_ladder}.

\subsection{Results and comparisons}

As for (Hpip)$_{2}$CuBr$_{4}$ we determined for (Hpip)$_{2}$CuCl$_{4}$ the exchange parameters from magnetization curves ($J_\perp= 3.42$ K, $J_\parallel= 1.34$ K, $g=2.06$). They compare favorably with other values found in the literature (Table~\ref{tab:exp}, \cite{HpipClBr_JMMM}). The two values of $J_\perp$ and $J_\parallel$ are significantly smaller than the ones for (Hpip)$_{2}$CuBr$_{4}$ confirming the important effects of the Cl--substitution on the exchange paths. Note that despite the change of the coupling constants the general structure of the phase diagram shown in Fig.~\ref{fig:phasediag} is still unchanged, and in particular the lower critical field $h_{c1}$ is still positive.

\begin{figure}
 \begin{center}
  \includegraphics[width=0.9\columnwidth]{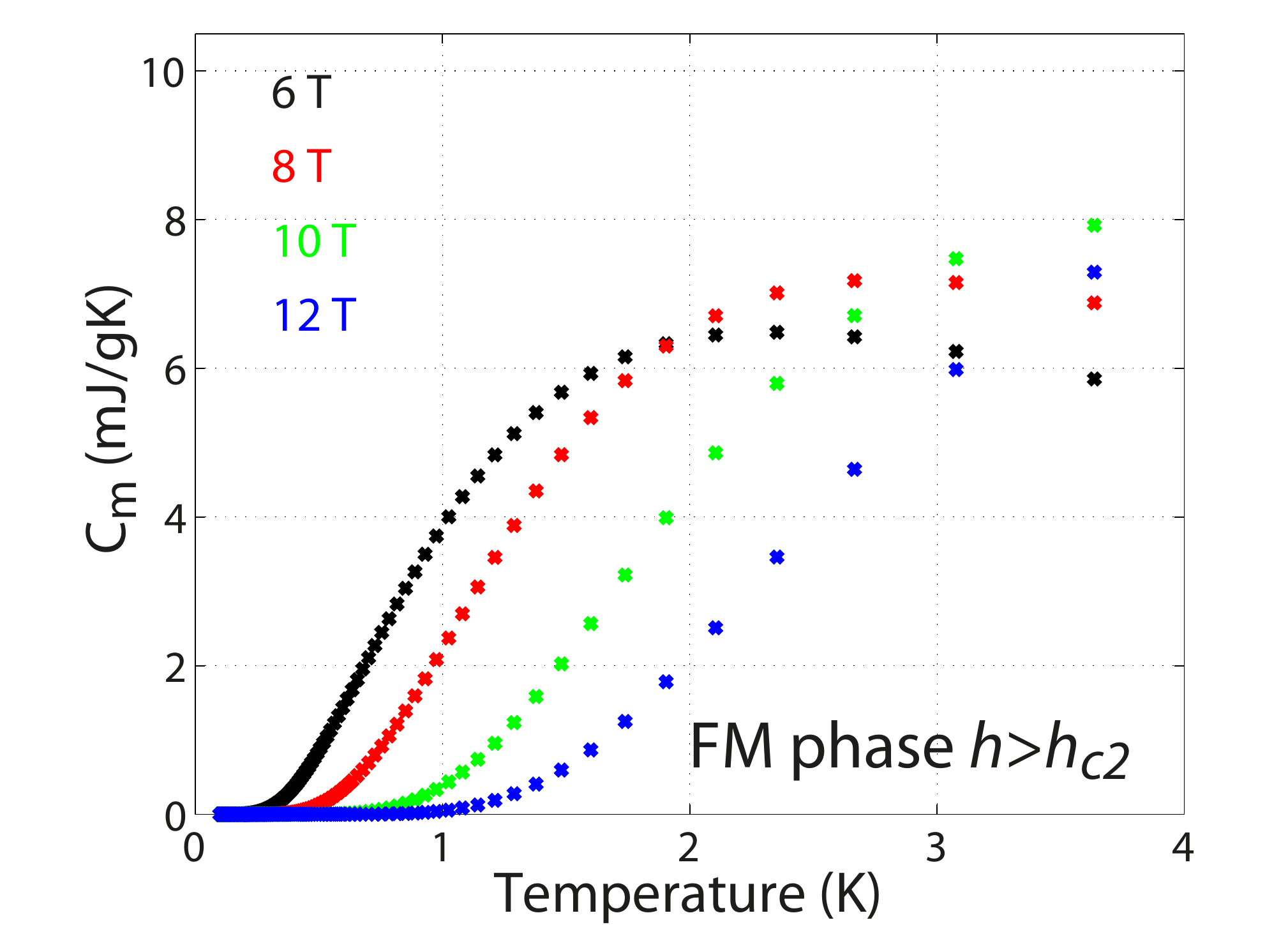}
 \end{center}
  \caption{\label{fig:aboveLL} Computed specific heat (see text) for the compound (Hpip)$_{2}$CuCl$_{4}$ for fields above the upper
  critical field $h_{c2}$ of Fig.~\ref{fig:phasediag} in the saturated phase (FM). The system is in a gapped phase as can be seen from the behavior of the specific
  heat. The gap decreases as one approaches $h_{c2}$ which signals the entrance in the gapless TLL regime.}
\end{figure}

The specific heat computed with the fixed exchange parameters is given below. Such calculations when compared with experiments can serve as a benchmark to confirm the Heisenberg form of the Hamiltonian for (Hpip)$_{2}$CuCl$_{4}$. We show first the high--field results $h>h_{c2}$ in Fig.~\ref{fig:aboveLL}. The presence of a gap is clearly visible on the higher field curves. As the magnetic field moves down the gap reduces as one approaches the TLL phase.

\begin{figure}
 \begin{center}
  \includegraphics[width=0.9\columnwidth]{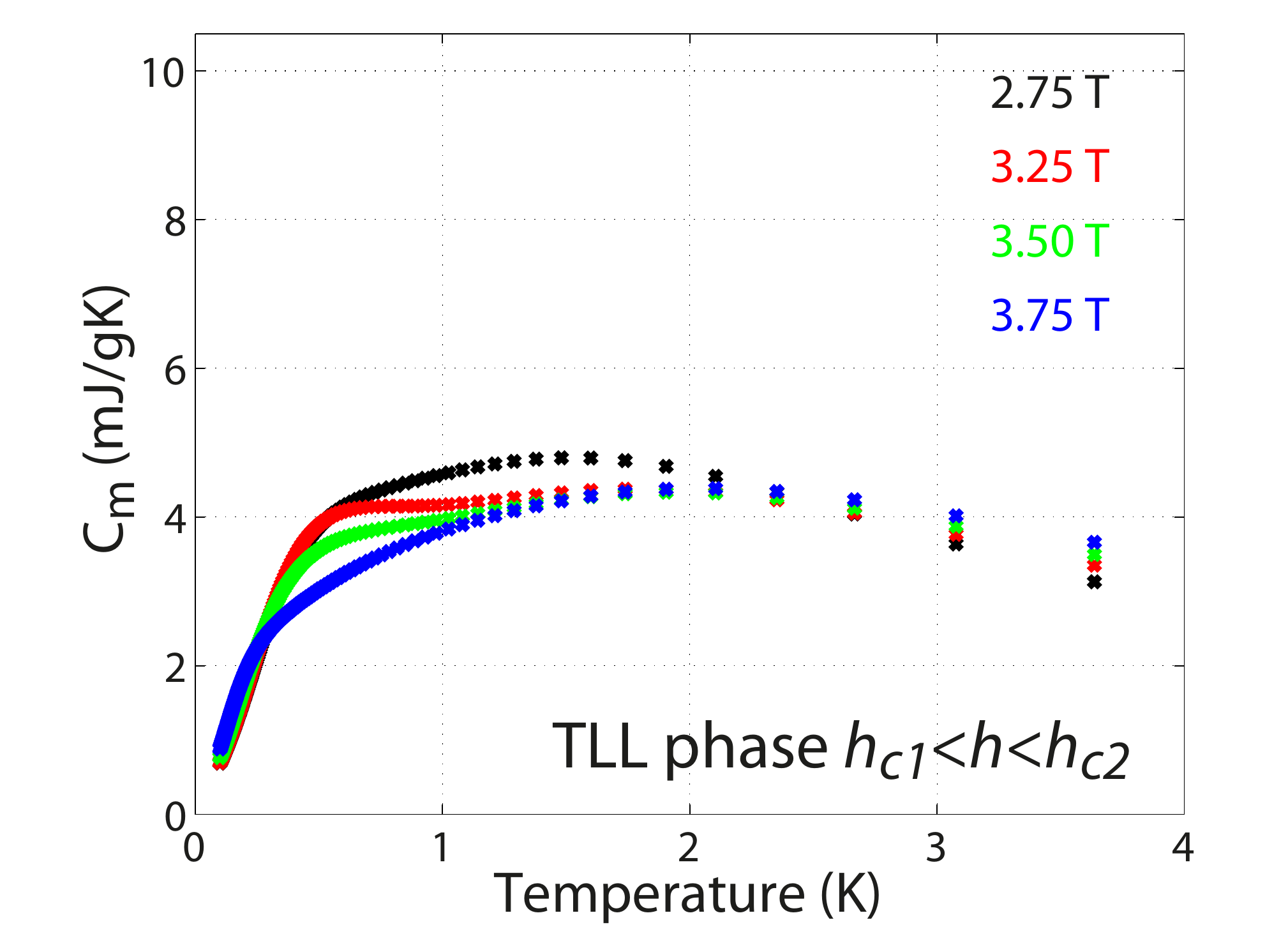}
 \end{center}
  \caption{\label{fig:insideLL} Computed specific heat (see text) for the compound (Hpip)$_{2}$CuCl$_{4}$ for fields within the TLL
   regime ($h_{c1} < h < h_{c2}$ as shown in Fig.~\ref{fig:phasediag}). The system is gapless and the low temperature specific heat
   is proportional to the temperature. The slope gives access to the inverse velocity of the spin excitations in the TLL phase (see text).}
\end{figure}

Data within the TLL phase is shown in Fig.~\ref{fig:insideLL} (for a comparison with the (Hpip)$_{2}$CuBr$_{4}$ see e.g. \cite{ruegg_thermo_ladder} and Fig.~6 of \cite{bouillot_dynamics_ladder_DMRG_long}). The gap has closed and the low energy part of the specific heat is now linear, as can be expected
in a TLL.

The slope of the specific heat is directly connected to the speed $u$ of the spin excitation in the TLL by $C_m(T) \propto T/u$ \cite{giamarchi_book_1d}. Note that the data shows a peak structure (most visible for the 3.25 T curve). This peak signals the coherence scale of the TLL (which can be also computed independently from the exchange constants and the filling of the band). It is less visible than for (Hpip)$_{2}$CuBr$_{4}$ because more masked by the contributions to the specific heat coming from the higher triplets in the spectrum.

\begin{figure}
 \begin{center}
  \includegraphics[width=0.9\columnwidth]{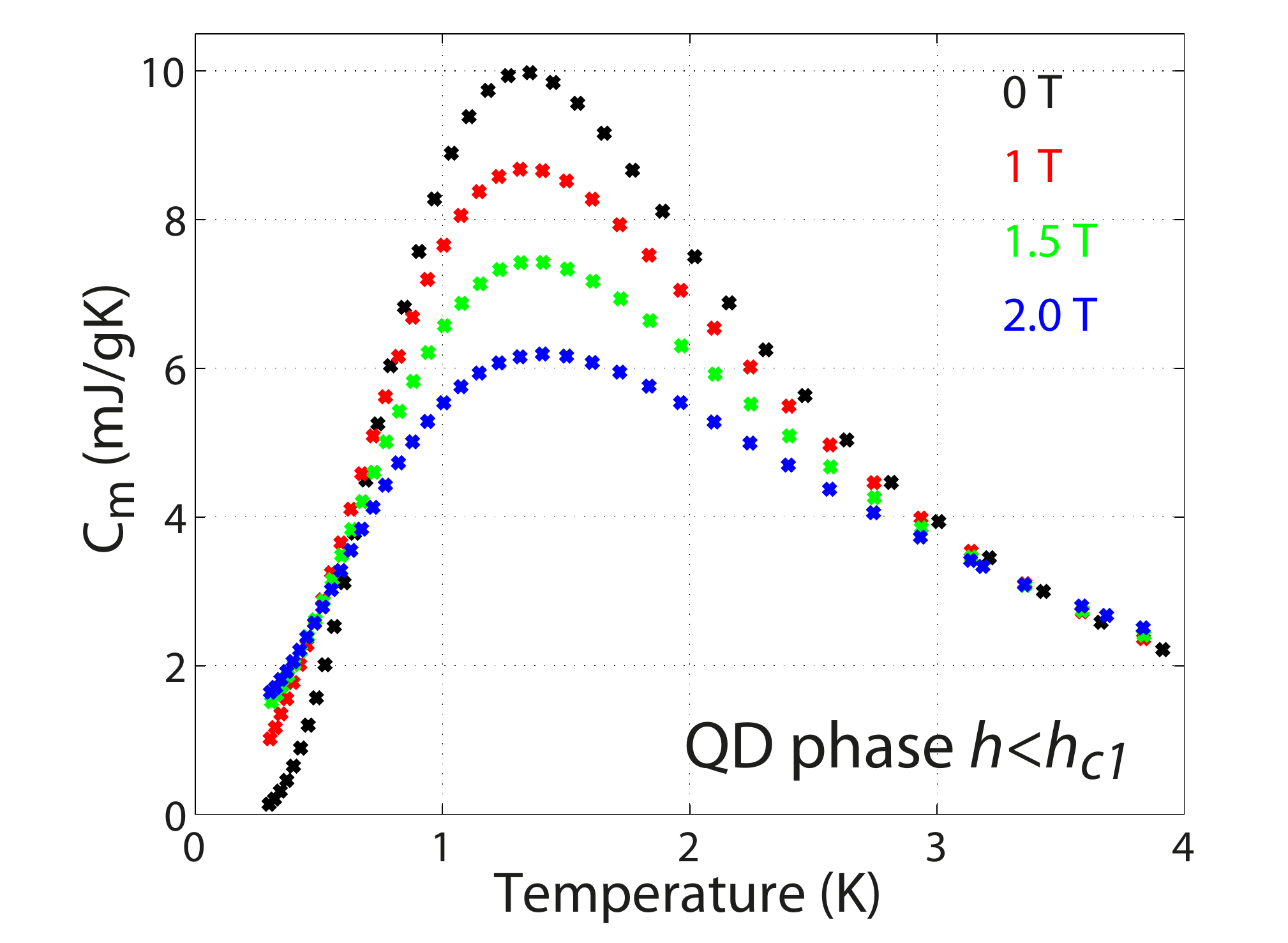}
 \end{center}
  \caption{\label{fig:belowLL} Computed specific heat (see text) for the compound (Hpip)$_{2}$CuCl$_{4}$ in the quantum disordered
  phase (QD) at low fields ($h < h_{c1}$ as shown in Fig.~\ref{fig:phasediag}). The system is gapped and the low temperature specific heat
   can be used to extract the spin gap.}
\end{figure}

Finally we show the low field computed data in Fig.~\ref{fig:belowLL}. As for the high field data the gap is again clearly visible and goes down as the field approaches $h_{c1}$.

The specific heat of (Hpip)$_{2}$CuCl$_{4}$ was measured on a purpose-built calorimeter at the Helmholtz Center Berlin on single crystals with a mass of 3.73mg between 0.3 K and 10 K using both quasi--adiabatic and relaxation techniques. Raw data for a small magnetic field $H=0.5$ T applied parallel to the crystallographic a-axis is shown in Fig. 4. The lattice contributes to the specific heat at low temperatures, but its signal is distinct from a clear maximum originating from magnetic excitations. The lattice contribution can be subtracted following a procedure that was applied successfully to (Hpip)$_{2}$CuBr$_{4}$ \cite{ruegg_thermo_ladder}.

\begin{figure}
 \begin{center}
  \includegraphics[width=0.9\columnwidth]{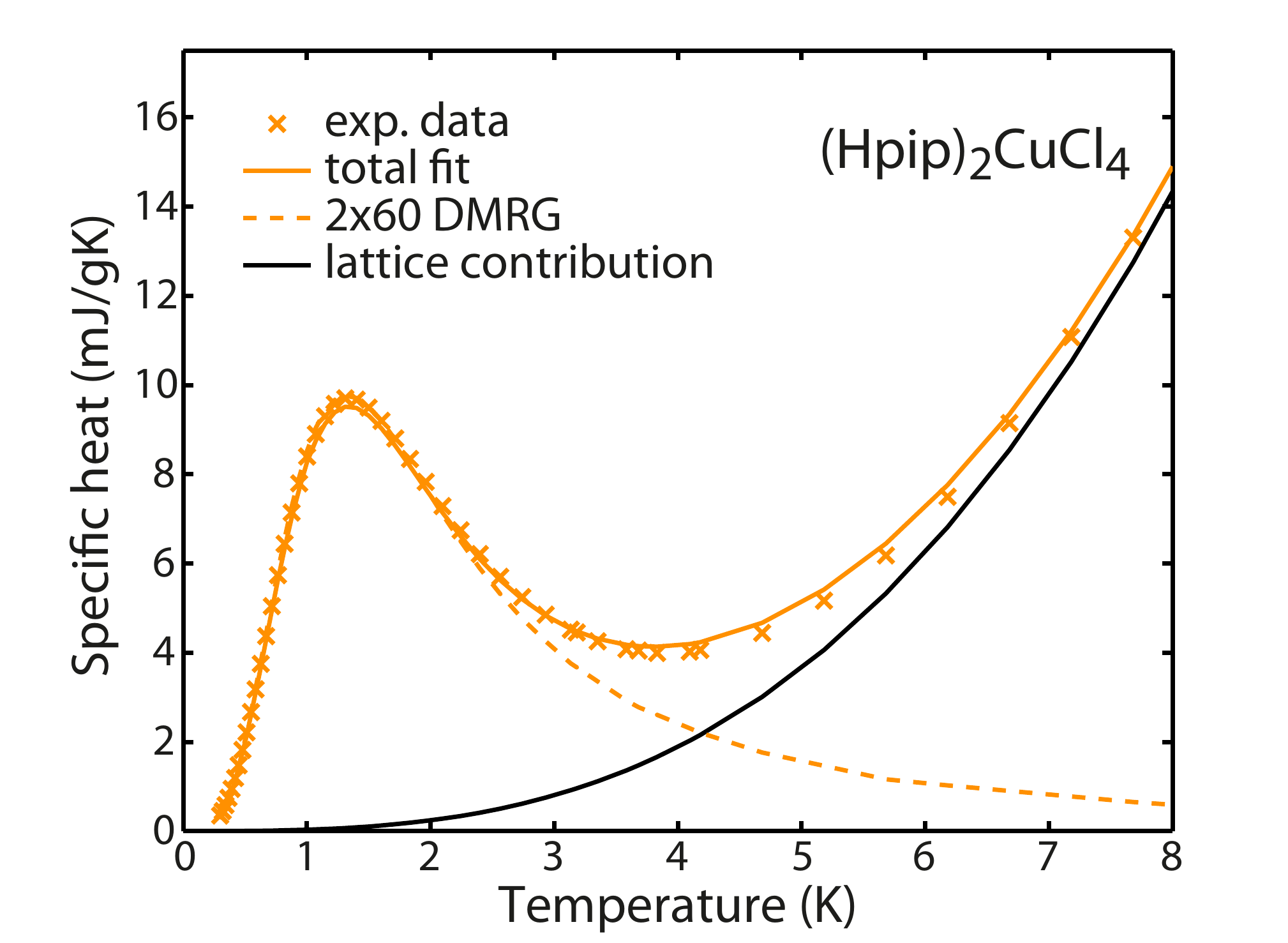}
 \end{center}
  \caption{\label{fig:comparison} Comparison between the measured and computed specific heat for (Hpip)$_{2}$CuCl$_{4}$ at $H=0.5$ T. Lines denote the spin (DMRG with 60 rungs (2x60 sites)) and lattice contributions. The remarkable agreement between theory and experiment vindicates the used form of the Heisenberg Hamiltonian and values of the exchange parameters. }
\end{figure}

The agreement, shown in Fig.~\ref{fig:comparison}, between theory and experiment is remarkable. It thus confirms, like for the parent bromide compound, that no major term is missing from the Hamiltonian (\ref{equ:spinladderhamiltonian}). It also confirms that the values of the exchange parameters that were determined by an independent method from the critical fields $h_{c1}$ and $h_{c2}$ are indeed accurate and more precise than in previous studies \cite{HpipClBr_JMMM}.

\section{Discussion and Conclusion} \label{sec:conclusion}

We have presented in this paper a determination of the Hamiltonian and of the exchange parameters that describe the compound (Hpip)$_{2}$CuCl$_{4}$.
The comparison between the measured specific heat in this compound and the calculations, using Density Matrix Renormalization Group calculations, show unambiguously that this compound is very well described by a spin ladder Heisenberg Hamiltonian.
The Hamiltonian and general phase diagram of (Hpip)$_{2}$CuCl$_{4}$ are similar to the ones of the parent compound (Hpip)$_{2}$CuBr$_{4}$, albeit with different exchange constants, and thus different critical fields as shown in Fig.~\ref{fig:phasediag}. Our study, which shows the similarity between the phase diagrams of the Cl-- and Br-- compounds thus pave the way to use the Cl-- substitution in (Hpip)$_{2}$CuBr$_{4}$ as a new route to realizing disordered TLLs. Indeed the substitution of a small concentration of Br-- by Cl-- will amount to locally change the exchange constants, or in the itinerant particle language to which the triplet excitation can be mapped, to realize a TLL with space dependent (disordered) hopping and chemical potential. Such a material should thus be a system of choice to tackle the effect of disorder in one dimensional systems, such as the existence and properties of the Bose glass. More generally this confirms the prominent role that the magnetic insulators can potentially play as quantum simulators.

\section*{Acknowledgements}
This work was supported in part by the Swiss NSF under MaNEP and Division II. 

\section*{Bibliography}

\end{document}